\documentstyle[aps]{revtex}

\input{tcilatex}

\begin{document}
\author{Mario Castagnino, Luis Lara}
\address{Instituto de F\'{i}sica de Rosario\\
Av. Pellegrini 250, Rosario, Argentina}
\author{Olimpia Lombardi}
\address{CONICET - Universidad de Buenos Aires. \\
Pu\'{a}n 470, Buenos Aires, Argentina}
\title{Time asymmetry as a consequence of the global properties of the universe.}
\maketitle

\begin{abstract}
A global definition of time-asymmetry is presented. Schulman's two arrows of
time model\cite{Schulman} is criticized.
\end{abstract}

\section{Introduction}

The problem of time-asymmetry, also known as the problem of the arrow of
time, can be summarized in the following question: {\it How an evident
time-asymmetry is possible if the laws of physics are time-reversal
invariant? }In fact, all the laws of physics are invariant under the
transformation $t\rightarrow -t$\footnote{%
There are two exceptions:
\par
i) The second law of thermodynamics: the entropy grows. But we use to
consider this ''law'' as an empirical fact that must be demonstrated from
more primitive and elementary laws.
\par
ii) Weak interactions. But they are so weak that it is difficult to see how
the asymmetry of the universe can be derived of these interactions.
Therefore, as it is usual in the literature, we do not address this problem
in this paper.}. Nevertheless, we have the psychological feeling that past
is different than future; moreover, there are clear time-asymmetric
phenomena, being the natural tendency from non-equilibrium to equilibrium
the most conspicuous example. Astonishing enough, the solution is contained
in the above italized question. Since there is a time-asymmetry that cannot
be explained by the time-reversal invariant laws ({\it equations}) of
physics, it should be explained by some time-asymmetric {\it initial
conditions.} But, at first sight, initial conditions are arbitrary;
therefore, it is impossible to formulate a physical law on initial
conditions. However, the initial conditions of any process are the result of
another process, in such a way that all processes in a connected universe
are coordinated in some way. Therefore, the reason of time-asymmetry is the
asymmetry of the universe, namely, a {\it \ global }reason. The aim of this
letter is to explain this fundamental fact in the clearest possible way, and
to discuss some recent criticisms of this global solution\cite{Schulman}.

Since Boltzmann's seminal work, many authors had the intuition that
time-asymmetry has a global origin. However, traditional discussions usually
define time-asymmetry in terms of entropy increasing. In this letter we will
reject the traditional entropic approach, following John Earman's \cite
{Earman} ''Time Direction Heresy'' according to which the arrow of time is
an intrinsic, geometrical feature of space-time: this geometrical approach
to the problem of the arrow of time has conceptual priority over the
entropic approach since the geometrical properties of the universe are more
basic than its thermodynamic properties.

\section{Time-orientability}

Earman \cite{Earman} and Gr\"{u}nbaum \cite{Grun} were the first authors who
emphasized the relevance of time-orientability to the problem of the arrow
of time. In fact, general relativity considers the universe as a
pseudo-Riemannian manifold that may be time-orientable or not. {\it \ }A
space-time is {\it time-orientable} if and only if there exists a {\it %
continuous }non-vanishing timelike vector field globally defined. By means
of this field, the set of all light semi-cones (lobes) of the manifold can
be splitted into two equivalence classes: $C_{+}$ and $C_{-}$. If space-time
were not time-orientable, the distinction between future lobes and past
lobes would not be univocally definable on a global level. On the other
hand, in a time-orientable space-time, if there were a time-reversal
non-invariant law {\it L}, defined in a {\it continuous} way all over the
manifold, that would allow us to choose one of the classes as the future one
(say $C_{+})$ and the other as the past one (say $C_{-})$: the law {\it L}
would be sufficient for defining the arrow of time for the whole universe
(namely, a future lobe $C_{+}(x)$ and a past lobe $C_{-}(x)$ at each point $%
x)$\footnote{%
Of course, a previous requirement is that a cosmic time could be defined in
the whole universe, namely, that the manifold would satisfy stable causality
condition \cite{HE} and that the monotonically increasing function $f$ could
be computed from the distance between two hypersurfaces of the corresponding
foliation along {\it any }curve orthogonal to the foliation.}. In fact, if
one lobe of the class $C_{+}$ were considered as the future lobe at a point $%
x$ and another lobe of the same class were considered as the past lobe at a
point $y$, then joining these two points with a continuous curve (because we
only consider connected universes) and propagating the lobe of $x$ towards $%
y $ (and vice-versa) would be sufficient for finding a point where the law 
{\it L} would be discontinuous, contrary to our supposition.

But, what is this global continuous time-reversal invariant law which allows
us to define past and future\footnote{%
See footnote 1.}? This is the essence of Matthews' criticism \cite{Matt} to
the relevance of time-orientability: since there are not continuous and
global time-reversal non-invariant laws of nature (but anyway the arrow of
time does exists), time-asymmetry is necessarily defined by local laws and,
then, it is just a local property; therefore, nothing rules out the
possibility that the arrow of time points to different senses in different
regions of space-time (see also Reichenbach \cite{Reichenbach}). What
Matthews has forgotten is that an asymmetric {\it physical fact} can be used
to define time-asymmetry instead of a physical law. Of course, it must be an 
{\it ubiquitous }physical fact, because it must be used to define the future
and the past lobes at all the points of the universe. We will show that the
time-asymmetry of the universe is this physical fact by arguing:

1.- That time-symmetric universes belong to a set of {\it measure zero} on
the space of all possible universes.

2.- That the global time-asymmetry of the universe can be used {\it locally }%
at each point $x$ to define the future and past lobes, $C_{+}(x)$ and $%
C_{-}(x).$

We will develop these points in the next two sections.

\section{The corkscrew factory theorem}

In his interesting book, Huw Price \cite{HP} emphasizes that time-reversal
invariance is not an obstacle to construct a time-asymmetric model of the
universe: a time-reversal invariant equation may have time-asymmetric
solutions\footnote{%
Of course, this ''loophole'' is not helpful when we are dealing with a
multiplicity of systems: for each time-asymmetric solution there may be
another time-asymmetric solution that is the temporal mirror image of the
first one. But when we are studying the whole universe, both solutions are
equivalent descriptions of one an the same universe.}. He illustrates this
point with the familiar analogy of a factory which produces equal numbers of
left-handed and right-handed corkscrews: the production as a whole is
completely unbiased, but each individual corkscrew is asymmetric. Price
argument shows the possibility of describing time-asymmetric universes by
means of time-reversal invariant laws. But, what is the reason to suppose
that time-asymmetric universes have high probability? We will demonstrate
that {\it time-asymmetric solutions of the universe equations have measure
zero en the corresponding phase space}.

Let us consider some model of universe equations. All known examples have
the following two properties (e.g. see \cite{Lara}, but there are many other
examples):

1.- They are time-reversal invariant, namely, invariant under the
transformation $t\rightarrow -t.$

2.- They are time-translation invariant, namely, invariant under the
transformation $t\rightarrow t+cont.$\footnote{%
We are referring to the equations that rule the behavior of the universe,
not to the particular solutions that normally do not have the time
translation symmetry.} (homogeneous time).

To fix the ideas let us consider the simplest example: a FRW universe with
radius $a$ and matter represented by a neutral scalar field $\phi .$ The
dynamical variables are $a,\stackrel{\bullet }{a},\phi ,\stackrel{\bullet }{%
\phi }$. They satisfy a Hamiltonian constraint $H(a,\stackrel{\bullet }{a}%
,\phi ,\stackrel{\bullet }{\phi })=0$ which reduces the dimension of phase
space from 4 to 3; then we can consider a phase space of variables $%
\stackrel{\bullet }{a},\phi ,\stackrel{\bullet }{\phi }.$ Let us now
consider a time-symmetric continuous\footnote{%
We will disregard non-continuous solutions since normally information does
not pass through discontinuities: we are only considering {\it connected}
universes where information can go from a point to any other time-like
connected point.} solution, for $a\geq 0$: there must be a time, that we can
take as the time origin $t=0$ (since the equations are time-translation
invariant), with respect to which $a(t)=a(-t);$ therefore $\stackrel{\bullet 
}{a}(0)=0$ (e.g., in a big bang-big crunch universe, there must be a
maximum; in a universe that begins and ends with a infinite radius there
must be a minimum at the bounce of the radius). This means that the radius
of the universe is symmetric with respect to $t=0$. But if we want to obtain
the complete time-symmetry of the universe, the field $\phi $ must be also
symmetric with respect to $t=0$. Since $\phi $ has not definite sign, we
have two possibilities: either $\phi (t)=\phi (-t)$ or $\phi (t)=-\phi (-t)$%
. Therefore at $t=0$ we have two possible boundary conditions: $(0,\phi ,0)$
or ($0,0,\stackrel{\bullet }{\phi })$ respectively. So the set of $t=0$
conditions that lead to time-symmetric solutions has dimension 1%
\mbox{$<$}
3, and the surface that contains time-symmetric solutions has dimension 2%
\mbox{$<$}
3. In the usual Lebesgue measure (or in any measure absolutely continuous
with respect to it) both the set of initial conditions that lead to
time-symmetric universes and the surface of time-asymmetric solutions have
measure zero. q. e. d. $.$

This theorem can be easily generalized to the case where $\phi $ has many
components, or to the case of many fields with many components where some of
these fields may be fluctuations of the metric. Since properties 1 and 2 are
also true in classical statistical mechanics, the theorem could be also
demonstrated in this case, and also in the quantum case, albeit some quantum
gravity problems like time definition\cite{PT}.

Let us now consider the coarse-grained version of the theorem. Let $%
\varepsilon $ be the size of the grain and, in order to compare measures,
let us consider that the phase space $\stackrel{\bullet }{a},\phi ,\stackrel{%
\bullet }{\phi }$ is a cube of volume $L^{3}.$ In this case, boundary
conditions $(0,\phi ,0)$ and ($0,0,\stackrel{\bullet }{\phi })$ will be
fuzzy and the volume of the set of time-symmetric initial conditions will
have measure $2\varepsilon ^{2}L.$ This magnitude can be compared with the
size of the phase space, obtaining the ratio $2\varepsilon
^{2}L/L^{3}=2(\varepsilon /L)^{2}.$ Of course, in the usual case $%
\varepsilon \ll L$; then, the measure of the set of points corresponding to
boundary conditions that lead to time-symmetric universes is extremely
smaller than the measure of the phase space. The same argument can be
applied to the set of time symmetric solutions with measure $\varepsilon
L^{2}$, where $\varepsilon L^{2}/L^{3}=\varepsilon /L<<1$ if $\varepsilon
\ll L.$ q.e.d.

This completes the first argument announced in Section II; let us now
develop the second argument.

\section{The Reichenbach-Davies diagram}

In his classical book about the arrow of time, Hans Reichenbach \cite
{Reichenbach} defines the future direction of time as the sense of the
entropy increasing of the majority of branch systems, that is, systems which
become isolated or quasi-isolated from the main system during certain
period. Paul Davies \cite{Davies} appeals to Reichenbach's notion, claiming
that branch systems emerge as the result of a chain or hierarchy of
branchings which expand out into wider and wider regions of the universe;
therefore ''the origin of the arrow of time refers back to the cosmological
initial conditions''.

On the basis of this idea, in previous papers we have introduced the
''Reichenbach-Davies diagram'' \cite{CastKL}, \cite{Peyresq}, \cite{Laciana}%
\footnote{%
At the quantum level it could be considered as the combination of all the
scattering processes within the universe. We have called the quantum version
of the ''Reichenbach-Davies'' diagram ''Reichenbach-Bohm'' diagram \cite
{Goslar}, \cite{Ordo}.}, where all the local processes which go from
non-equilibrium to equilibrium are connected in such a way that the
''output'' of a process is the ''input'' of another one: the energy provided
by a process relaxing to equilibrium serves to drive another process to
non-equilibrium. This ''cascade'' of processes define a global energy flux
which, if traced back, owes its origin to the initial global instability
that is the source of all the energy of the universe\footnote{%
The initial instability is studied in \cite{Aqui} and bibliography therein.}%
. On the other hand, the sense of the flux on the time-orientable space-time
defines a global {\it time-orientation}: the incoming flux defines the lobe $%
C_{-}(x)\in C_{-}$ at each point $x$, the outgoing flux defines the lobe $%
C_{+}(x)\in C_{+}$ at $x$, and all the lobes of class $C_{-}$ point towards
the initial instability. In this way, the global time-asymmetry of the
universe defines the local time-asymmetry in each one of its points.

In summary, the Reichenbach-Davies diagram shows how global time-asymmetry
is related with local time-asymmetry, that is, how the different ''arrows of
time'' (cosmological, thermodynamic, quantum, electromagnetic, etc.) are
coordinated \cite{Peyresq}. The global energy flux is the ubiquitous
phenomenon that connects all the processes of the universe. If two sections
of the universe are not connected by this flux, then they are completely
isolated from each other: each section can be considered as a universe by
itself, and within each one of them the global flux is unique. These
considerations are particularly relevant for evaluating Schulman's argument.

\section{Criticism to Schulman\'{}s argument}

Schulman\cite{Schulman} exhibits a model in which two weakly coupled systems
maintain opposite running thermodynamic arrows of time. From this model, he
concludes that regions of opposite running arrows of time at stellar
distances from us are possible. This possibility would be a counter-example
to our position: Schulman's model would show that a universe consisting in
two weakly coupled sub-universes A and B can have two regional arrows of
time pointing to opposite senses.

Even though Schulman's argument sounds convincing at first sight, it becomes
implausible when analyzed form a cosmological viewpoint. In Schulman's
proposal, the low entropy extremities of the sub-universes A and B are
opposed, and both sub-universes evolve towards equilibrium in opposed time
senses. Let us consider two cases:

1.- The sub-universe A is bigger than the sub-universe B\footnote{%
To fix the ideas we can say ''externally bigger''. The cases ''more or less
bigger '' or ''almost symmetric''can be included in the coarse-grained
version of the cork-screw theorem, since in these cases there is only a
small difference with a symmetric model. Then these solutions have small
measure.} (this situation is not considered by Schulman). If
time-orientation is defined by entropy increasing, in this case the
time-orientation of the whole universe A$\cup $B will agree with the
time-orientation of A, and B will go from equilibrium to non-equilibrium.
Nevertheless, the behavior of B is neither strange not unnatural: since
there is a flux of energy which, according to the time-orientation adopted,
must be considered as a flux from A to B, then we can consider that it is
such energy what takes the sub-universe B out of equilibrium. In other
words, the decreasing entropy of the {\it open} sub-universe B has the same
explanation as the decreasing entropy in the usual open systems that we find
in our everyday life.

2.- The sub-universe A is equal to B (the situation studied by Schulman,
where A and B are identical). In this case, the universe A$\cup $B is
perfectly time-symmetric. But, as the argument of Section III has shown,
time-symmetry has vanishing measure: it requires an overwhelmingly
improbable fine-tuning of all the state variables of the universe.

But even in the time-symmetric case, it is not admissible to suppose that
the sub-universes A and B have opposite time-orientations. Schulman's
cosmological model is a time-orientable universe. In a time-orientable
manifold, continuous timelike transport has conceptual priority over any
method of defining time-orientation. In other words, Schulman's universe has
a light-cone structure such that, if we continuously transport a future
pointing vector from the point $x\in $A along some curve to the point $y\in $%
B, the transported vector will fall into the future lobe $C_{+}(y)$. This
means that A's future cannot be different than B's future: there is an only
future for the whole manifold, defined by its light-cone structure.

However, who prefers to insist on the attempt to use entropy increasing for
defining time-orientation could appeal to the following strategy: to define
the future sense of time as the sense of the entropy increasing, for
instance, in the sub-universe A, and then to establish the time-orientation
in the sub-universe B by means of continuous timelike transport. But who
adopts this strategy is committed to explain why future is the sense of
entropy increasing in one region of the universe but not in the other: why
the entropy definition works in one region of the universe but not in all of
them.

These considerations lead us to our starting point: the problem of the arrow
of time should be addressed from a global perspective, taking into account
the geometrical properties of space-time.

\section{Conclusion.}

There is never a last word in physics. But (provisionally) we can conclude
that the global definition of the arrow of time has no serious faults and,
therefore, it can be used as a solid basis for studying other problems
related with the time asymmetry of the universe and its sub-systems.

\end{document}